\documentclass[aps, pop, onecolumn, superscriptaddress, nofootinbib, preprint]{revtex4-2}
\usepackage{amsmath,amssymb,graphicx,color}
\usepackage[FIGTOPCAP]{subfigure}
\usepackage{bm}

\begin{document}

\title{Generalization of the Rosenbluth-Putvinski avalanche growth rate to the drift kinetic regime}

\author{Christopher J. McDevitt}
\email{cmcdevitt@ufl.edu}
\affiliation{Nuclear Engineering Program, Department of Materials Science and Engineering, University of Florida, Gainesville, FL 32611, United States of America}
\author{Jonathan S. Arnaud}
\affiliation{Theoretical Division, Los Alamos National Laboratory, Los Alamos, NM 87545, United States of America}

\date{\today}

\begin{abstract}

A new collisionless detrapping mechanism is shown to modify the rate of runaway electron (RE) generation by as much as a factor of two for strong inductive electric fields. This detrapping mechanism results from the magnetic well inherent to tokamak geometry being tilted by the strong inductive electric fields present during tokamak disruptions. The impact of this collisionless detrapping mechanism is found to be described by a single dimensionless parameter, whose identification enables the transition between distinct regimes of RE generation to be characterized. Using this composite parameter, it is shown that modifications to the avalanche growth rate by tokamak geometry across a broad range of plasma parameters collapse onto a one-dimensional curve that can be accurately fit with a two parameter function. The resulting analytical formula extends the classic Rosenbluth-Putvinski avalanche growth rate expression beyond the bounce-averaged limit into RE regimes where collisionless detrapping strongly modifies RE generation rates.

\end{abstract}

\maketitle

\newpage

\section{Introduction}
\label{sec:I}

Runaway electrons (RE), despite their importance to magnetic fusion plasmas~\cite{Knoepfel:1979, breizman2019physics}, atmospheric plasmas~\cite{wilson1925acceleration, Gurevich:1992} and solar flares~\cite{holman1985acceleration}, remain incompletely explained. Such energetic electron populations, while making up a small fraction of the total electron population, can carry the majority of the plasma current during a tokamak disruption~\cite{Hender:2007, nardon2017progress, ferraro2019two, hoelzl2021jorek}, can initiate terrestrial gamma-ray flashes in the earth's atmosphere~\cite{fishman1994discovery}, and provide a potential mechanism for hard X-ray generation in solar flares~\cite{benz2017flare}. An outstanding challenge to their description in a tokamak plasma is that the efficiency of RE generation can be substantially modified by the magnetic geometry present in a tokamak device. Such an effect, first described in the classic work of Rosenbluth and Putvinski (RP97)~\cite{Rosenbluth:1997} in the context of the RE avalanche~\cite{Sokolov:1979, Jayakumar:1993}, follows due to magnetically trapped electrons being alternately accelerated and decelerated by a DC electric field. After a complete orbit, the electrons receive no net energy gain, thus leading to a marked reduction in the efficiency of RE generation~\cite{Chiu:1998, eriksson2003simulation, Nilsson:2015, mcdevitt2019avalanche, guo2019toroidal, hoppe2021dream}.

This physical picture is amended when a strong inductive electric field is present. Evidence for such a modification was first numerically observed in Ref. \cite{mcdevitt2019runaway}. Here, it was noted that for the large inductive electric fields typical of a tokamak disruption, the threshold energy for RE generation is sufficiently low that electrons are no longer in the asymptotically small collisionality regime, violating a key assumption in the classic work by RP97~\cite{Rosenbluth:1997}. Subsequent work in Ref. \cite{arnaud2024impact} provided a more comprehensive treatment of disruption scenarios, thus quantifying deviations from RP97's theory across the full disruption landscape. In this paper, we elucidate the physical mechanism underlying deviations from RP97's theory in the strong electric field regime, identify a single dimensionless parameter governing the transition between two limiting RE generation regimes, and use it to construct a new analytic expression for the RE avalanche growth rate spanning both limits.

To illustrate how strong electric fields modify the physics of RE generation we begin with a drift kinetic~\cite{Cary:2009} description of relativistic electron evolution in a tokamak with nested magnetic flux surfaces. We assume the device is large enough that cross-field drifts are negligible~\cite{nilsson2015trapped, mcdevitt2018spatial}, so orbits remain confined to a flux surface. The characteristic equations are taken to be:
\begin{subequations}
\label{eq:PF1}
\begin{equation}
\frac{dp}{dt} = - \frac{\xi E_0}{1+\varepsilon_r \cos \theta}  - \frac{\gamma^2}{p^2}
, \label{eq:PF1a}
\end{equation}
\begin{equation}
\frac{d\xi}{dt} = - \frac{\left( 1 - \xi^2 \right)}{1+\varepsilon_r \cos \theta} \left( \frac{E_0}{p} + \frac{1}{2} v \tau_0 \varepsilon_r \sin \theta \right)
, \label{eq:PF1b}
\end{equation}
\begin{equation}
\frac{d\theta}{dt} = v \xi \tau_0
. \label{eq:PF1c}
\end{equation}
\end{subequations}
Time is normalized to the relativistic collision time $\tau_c \equiv 4\pi \varepsilon^2_0 m^2_e c^3/ \left( n_e e^4 \ln \Lambda \right)$, momentum to $p \to p/(m_e c)$ and speed to $v\to v/c$, where $n_e$ is the free electron density and $\ln \Lambda$ is the Coulomb logarithm. We've assumed circular flux surfaces such that $R/R_0 = 1 + \varepsilon_r \cos \theta$, $b_g \equiv B/B_\varphi = \sqrt{1+\left( \varepsilon_r/q\right)^2}$, $\varepsilon_r = \left(r/a\right) \varepsilon$, $\varepsilon=a/R_0$ and $q=rB_0/ \left( R_0 B_\theta \left( r\right)\right)$. The toroidal electric field is $E_\varphi = \left( R_0/R\right) E_1$, and we defined $E_0 \equiv E_1/\left( E_c b_g\right)$, $\tau_0 \equiv \tau_c c / \left( qR_0 b_g\right)$, where $E_c=m_ec/ \left( e\tau_c\right)$ is the Connor-Hastie field~\cite{Connor:1975}, $\gamma^2=1+p^2$ and $\xi\equiv p_\Vert/p$. Equation (\ref{eq:PF1a}) describes the balance between electric field acceleration and drag. Equation (\ref{eq:PF1b}) contains the pinch by the electric field toward the RE direction $\xi=-1$ and the magnetic mirror term, with Eq. (\ref{eq:PF1c}) describing parallel streaming. Pitch-angle scattering is included using the Monte Carlo operator described in Ref. \cite{Boozer:1981}, with $\tau_c \nu_D = \left( \gamma/p^3\right) \left( 1 + Z_{eff}\right)$ and $Z_{eff}$ is the effective charge. 


\begin{figure}
\centering
\begin{centering}
\includegraphics[scale=0.75]{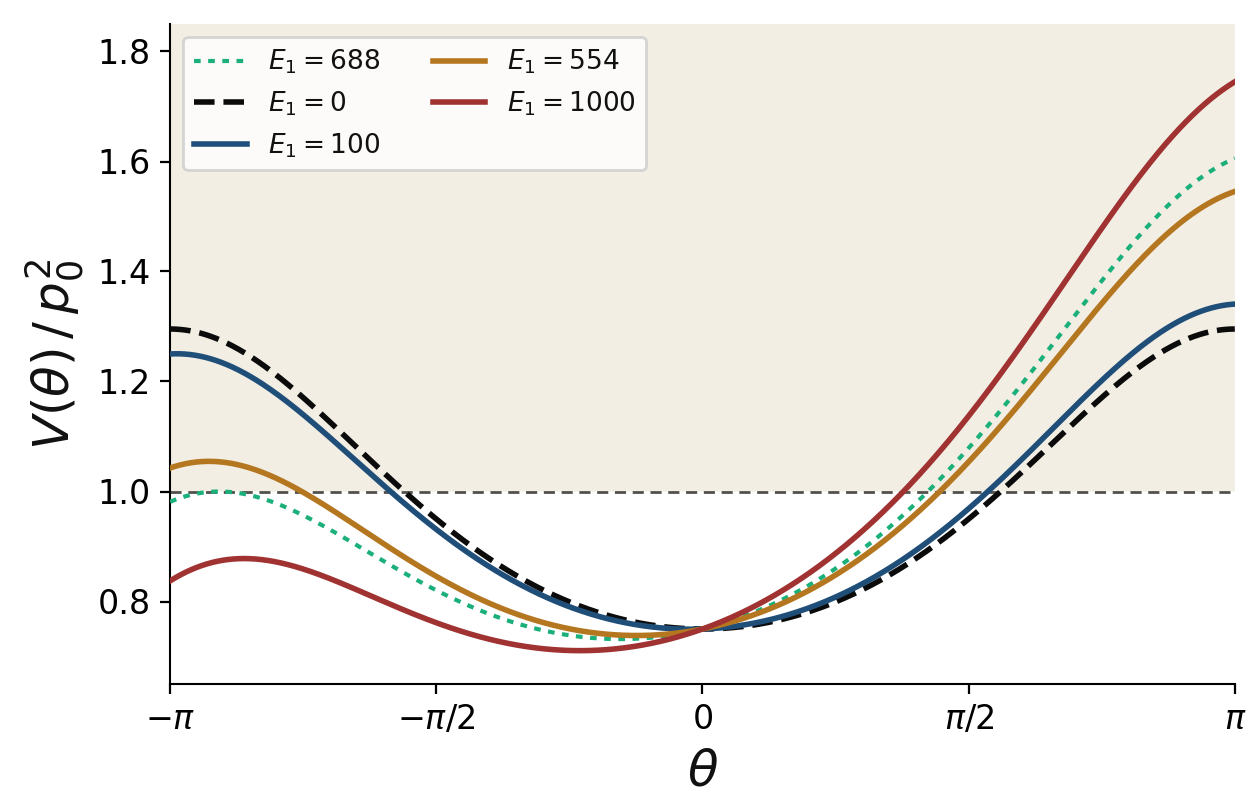}
\par\end{centering}
\caption{\label{fig:TiltWell} The magnetic well tilted by an inductive electric field [Eq. (\ref{eq:Well1})]. The parameters were taken to be $\left( c\tau_c/a\right) = 10^6$, $q=2$, $\varepsilon = 1/3$, $r/a=0.8$, $p_0 = 0.3$ and $\xi_0=0.5$. }
\end{figure}

To demonstrate the modification to the magnetic trapping criterion by a strong inductive electric field we will initially consider the collisionless limit. 
Noting that for a finite loop voltage, the energy of an electron is not conserved, Eqs. (\ref{eq:PF1a}) and (\ref{eq:PF1c}) can be combined to describe the variation of the electron's kinetic energy with poloidal angle, i.e.
\begin{equation}
\frac{d\gamma}{d \theta} = \frac{d\gamma/dt}{d\theta/dt} = - \frac{E_0}{\tau_0} \frac{1}{1+\varepsilon_r \cos \theta}
, \label{eq:dgdtheta}
\end{equation}
yielding
\begin{equation}
\gamma \left( \theta \right) = \gamma_0 - \frac{E_0}{\tau_0} I \left( \theta\right)
, \label{eq:gamma}
\end{equation}
where $d\gamma/dt = v dp/dt$, $\gamma_0$ is the Lorentz factor at $\theta=0$ and $I \left( \theta \right) \equiv \int^\theta_0 d \theta^\prime/\left( 1+\varepsilon_r \cos \theta^\prime \right)$. 
A trapping criterion can be derived by writing the relativistic momentum as $p^2= p^2_\Vert + p^2_\perp$ and substituting Eq. (\ref{eq:gamma}) in for $p^2=\gamma^2-1$, yielding an expression for the parallel momentum
\begin{equation}
p^2_\Vert = \gamma^2 - 1 - p^2_\perp = \left[ \gamma_0 - \frac{E_0}{\tau_0} I \left( \theta\right) \right]^2 - 1 - 2\mu B \left( \theta \right)
, \label{eq:pVert}
\end{equation}
where $\mu \equiv p^2_\perp / \left( 2B\right)$ is the magnetic moment. Defining the magnetic well
\begin{equation}
V \left( \theta \right) \equiv 2 \mu B \left( \theta \right) - \left[ \gamma^2 \left( \theta \right) - \gamma^2_0 \right]
, \label{eq:Well1}
\end{equation}
and noting the bounce points will occur when $p_\Vert = 0$, allows Eq. (\ref{eq:pVert}) to be rewritten as
\begin{equation}
0 = p^2_0 - V \left( \theta \right)
. \label{eq:Well2}
\end{equation}
Magnetic trapping requires $V \left( \theta \right) > p^2_0$. For $E_0=0$, $\gamma$ will be constant yielding $V \left( \theta \right) = 2\mu B \left( \theta \right)$, implying a symmetric magnetic well centered about the weak field side ($\theta=0$).
If a finite inductive electric field is present, $\gamma$ evolves with $\theta$ according to Eq. (\ref{eq:gamma}), such that the magnetic well is tilted as shown in Fig. \ref{fig:TiltWell}.
As indicated by the dotted green curve in Fig. \ref{fig:TiltWell} a threshold electric field emerges where $V \left( -\left|\theta \right|\right) < p^2_0$, such that the magnetic well is no longer able to confine the electron.


\begin{figure*}
\begin{centering}
\subfigure[]{\includegraphics[scale=0.22]{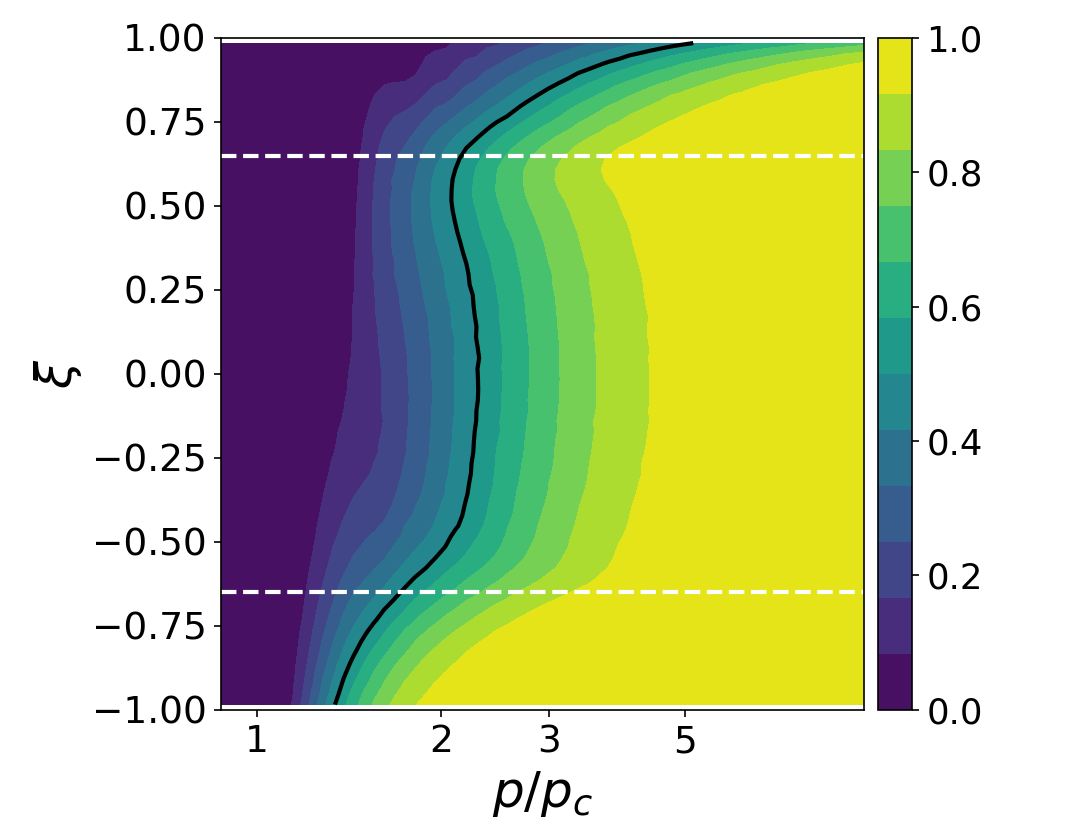}}%
\subfigure[]{\includegraphics[scale=0.23]{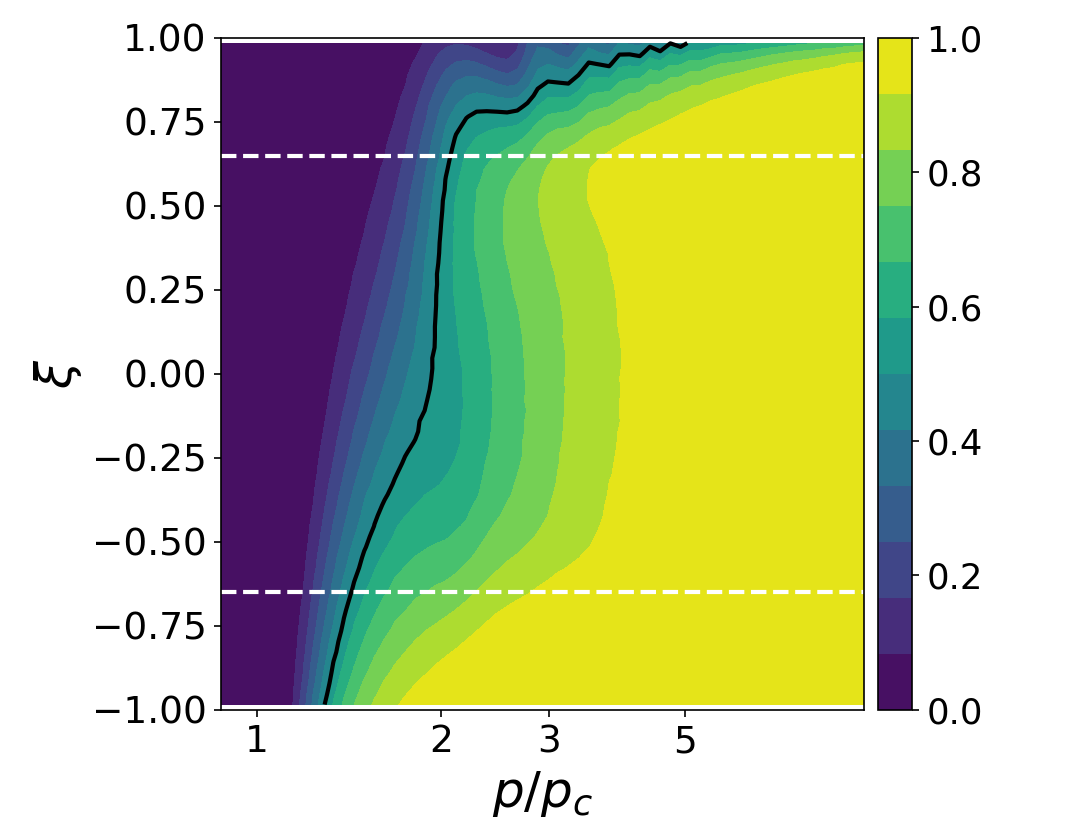}}%
\subfigure[]{\includegraphics[scale=0.23]{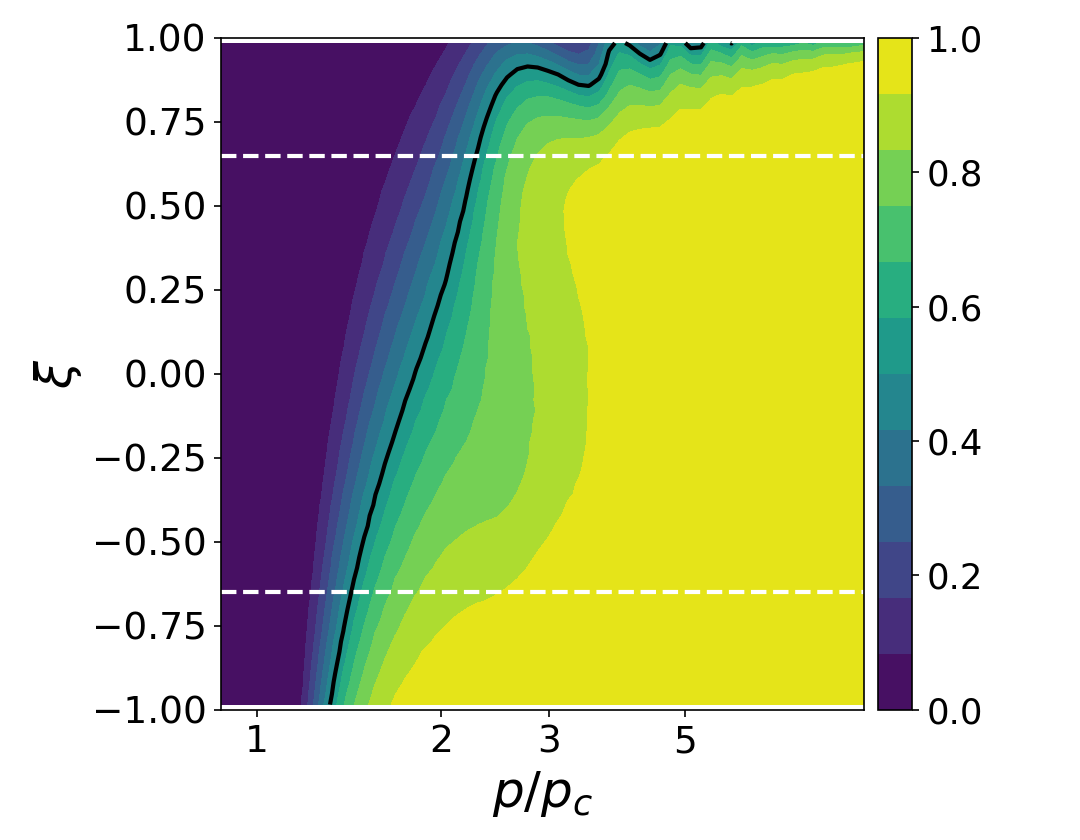}}%
\subfigure[]{\includegraphics[scale=0.23]{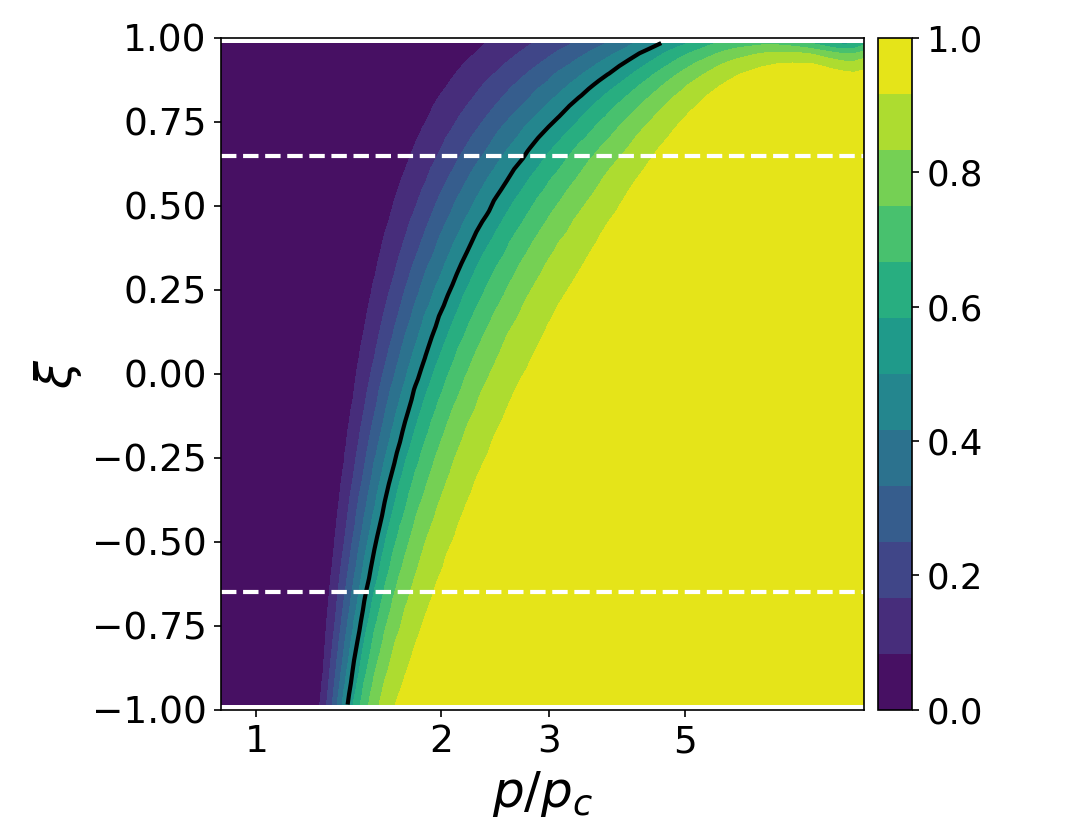}}
\subfigure[]{\includegraphics[scale=0.23]{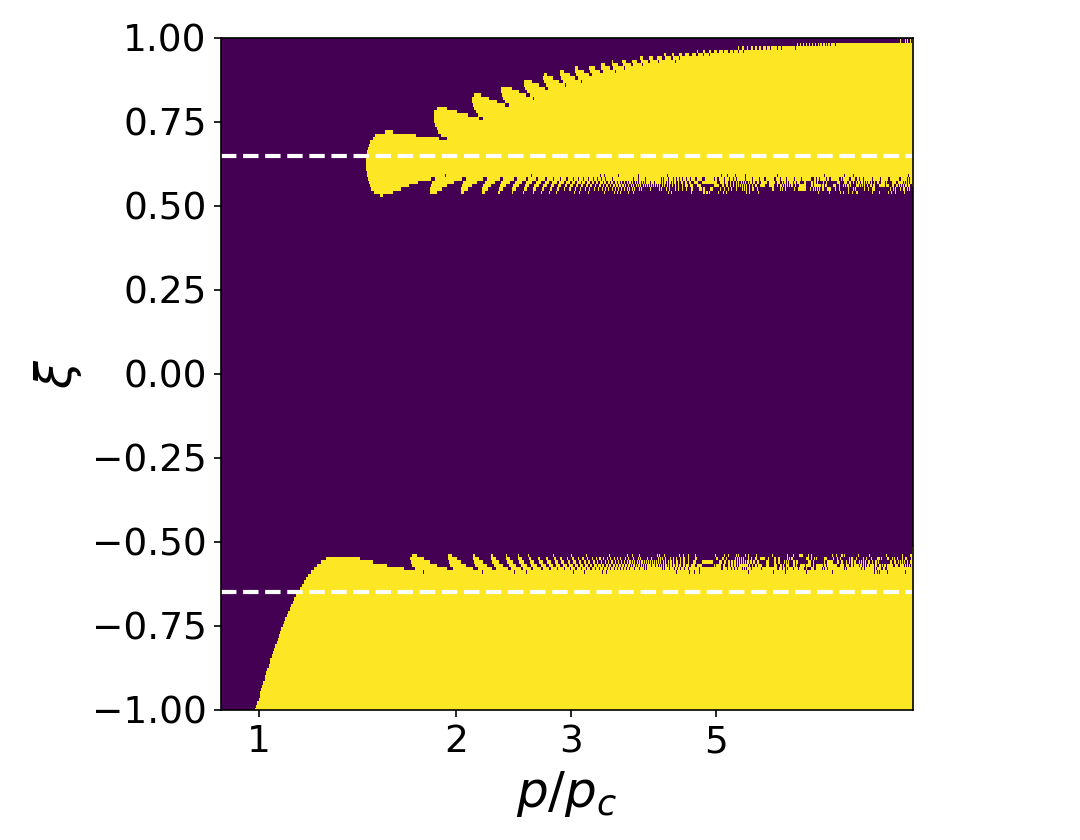}}%
\subfigure[]{\includegraphics[scale=0.23]{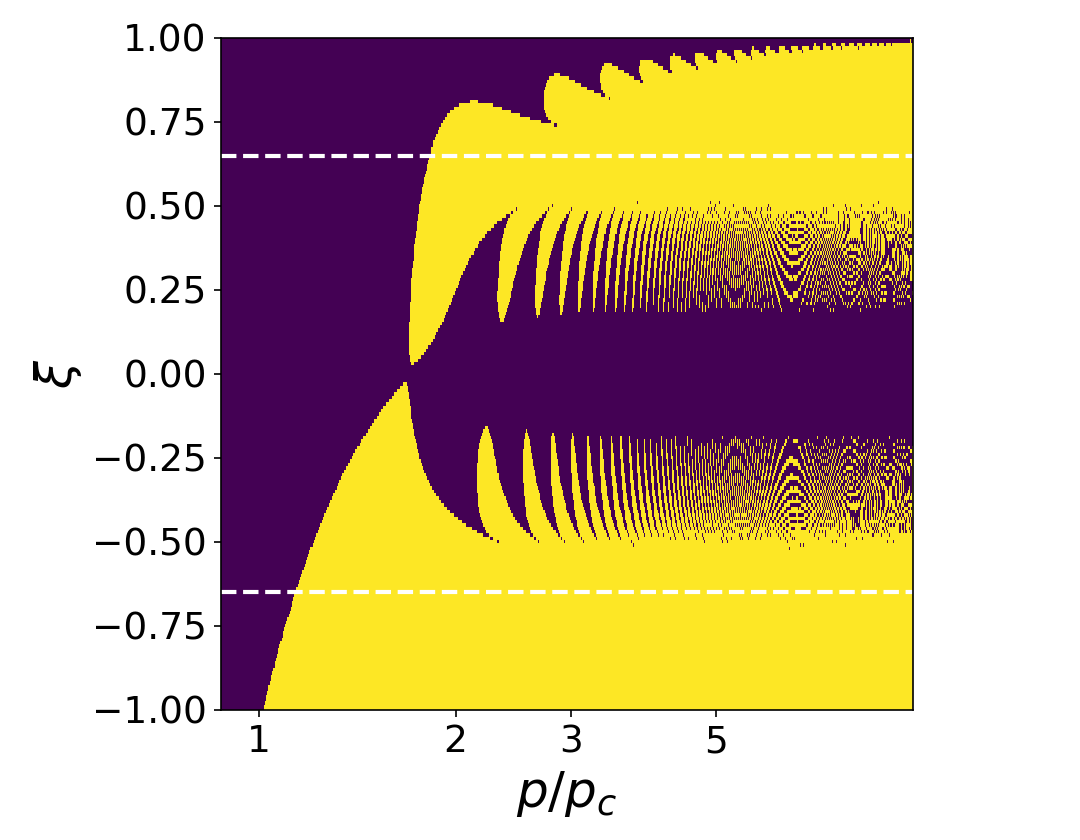}}%
\subfigure[]{\includegraphics[scale=0.23]{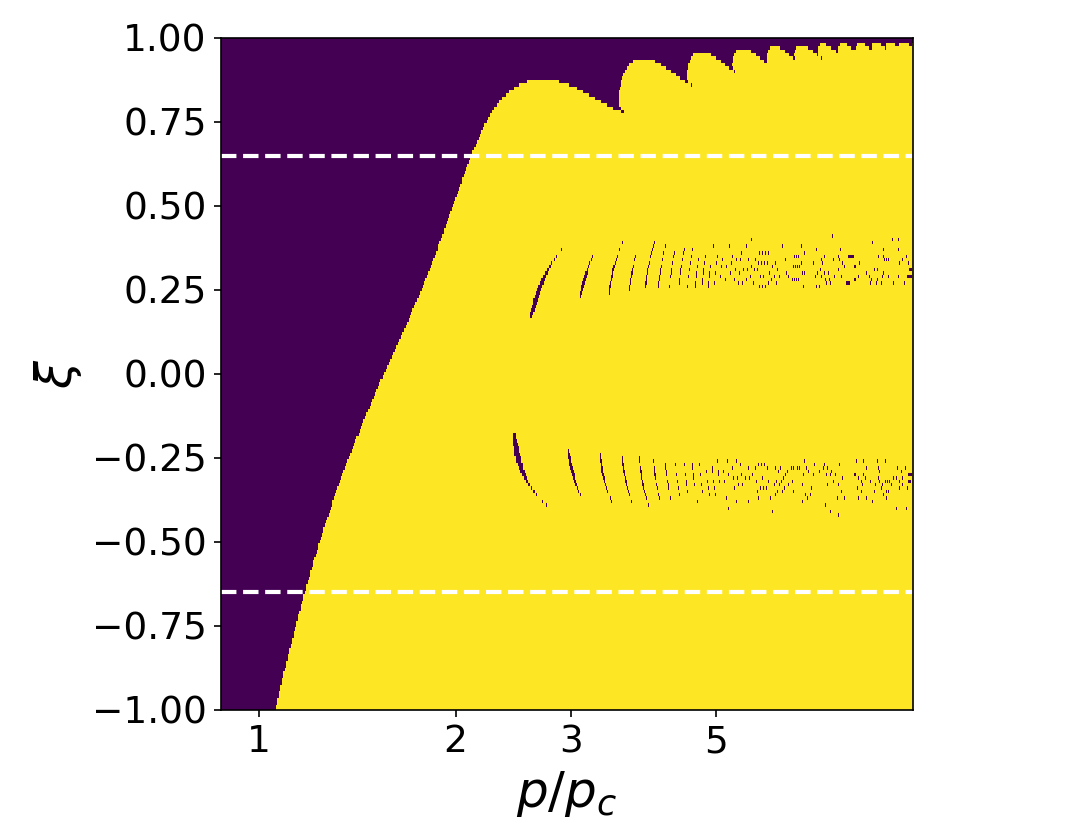}}%
\subfigure[]{\includegraphics[scale=0.23]{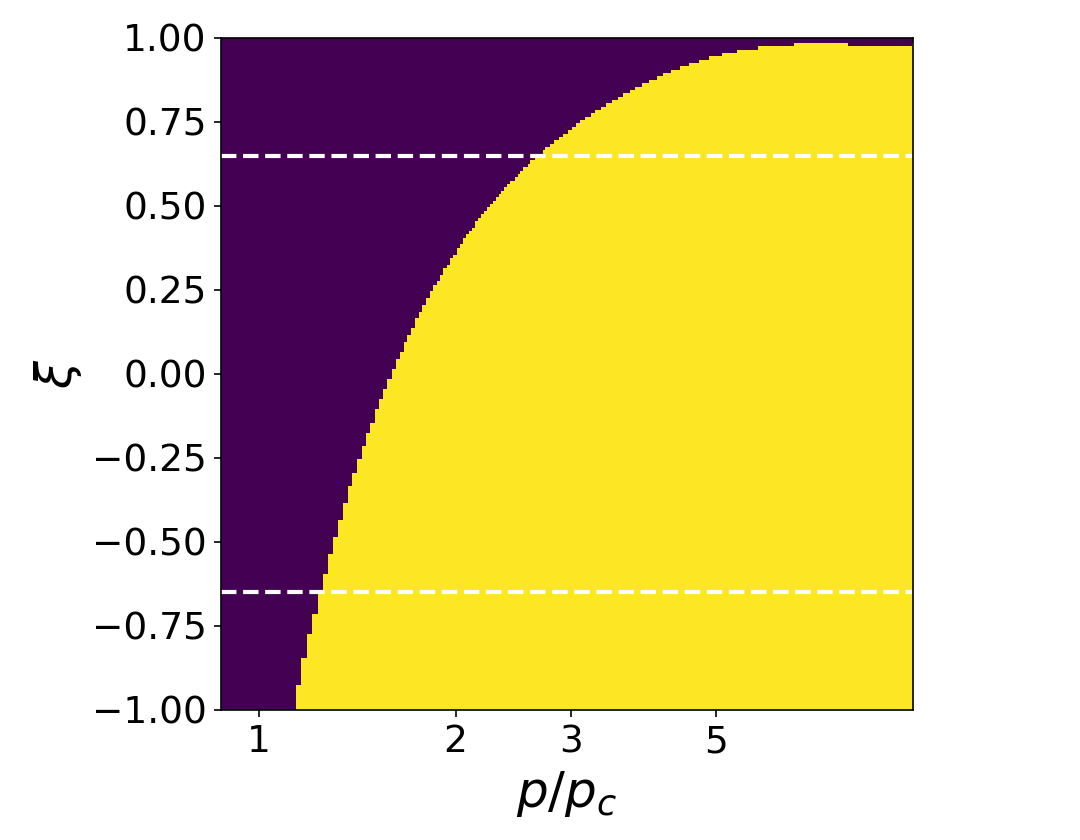}}
\par\end{centering}
\caption{\label{fig:RPF1} Cross cut of the RPF at $\theta=0$ for different electric field strengths. Panel (a) is for $E_1=100$, panel (b) is for $E_1=200$, panel (c) is for $E_1=300$ and panel (d) is for $E_1=1000$. Panels (e-h) neglect pitch-angle scattering, yielding fate maps with yellow indicating one and purple zero. The other parameters were set to $\left( c\tau_c/a\right) =10^6$, $r/a = 0.8$,  $a/R_0=1/3$, $Z_{eff} = 2$, $q = 2$ and $p_{max} = 10\,p_c$. The dashed lines indicate the trapped-passing boundary and the solid black curves in the first row indicates the $P=0.5$ contour.}
\end{figure*}

An approximate criterion for the opening of the magnetic well can be derived by expanding Eq. (\ref{eq:Well1}) for $E_0/\tau_0 \ll 1$, yielding
\begin{align}
V \left( \theta \right) &\approx 2\mu B \left( \theta \right) - \left[ \gamma^2_0 - 2 \frac{E_0}{\tau_0} \gamma_0 I \left( \theta \right) - \gamma^2_0 \right] = p^2_0 \left( 1- \xi^2_0 \right) \frac{1+\varepsilon_r}{1+\varepsilon_r \cos \theta} + 2 \frac{E_0}{\tau_0} \gamma_0 I \left( \theta \right)
, \label{eq:ApproxV}
\end{align}
where we inserted $\mu = p^2_0 \left( 1 - \xi^2_0\right)/ \left( 2 B \left( 0 \right) \right)$ and noted $B \left( \theta \right)/B \left( 0 \right) = \left( 1 + \varepsilon_r \right) / \left( 1 + \varepsilon_r \cos \theta \right)$. Substituting Eq. (\ref{eq:ApproxV}) into (\ref{eq:Well2}), and requiring $V \left( \theta \right)$ to be greater than $p^2_0$, yields the trapping criterion
\begin{equation}
\max_{\theta_b} \frac{1}{\left| I \left( \theta_b \right) \right|} \frac{1}{\varepsilon_r} \left[ \left( 1 - \xi^2_0 \right) \frac{1+\varepsilon_r}{1+\varepsilon_r \cos \theta_b} - 1 \right] \geq 2 \frac{\gamma_0}{p^2_0} \frac{E_0}{\varepsilon_r\tau_0}
, \label{eq:TrapCrit}
\end{equation}
where we have defined the bounce angle $\theta_b = - \left| \theta_b\right|$ to be negative and the left hand side should be evaluated at the $\theta_b$ that maximizes this function.
The depth of the well is set by the value of $\varepsilon_r$. This dependence can be extracted by first noting that trapped electrons will have $\xi^2_0 < \xi^2_t \approx 2 \varepsilon_r$, and expanding Eq. (\ref{eq:TrapCrit}) to first order in $\varepsilon_r$, yields
\begin{equation}
\max_{\theta_b} \frac{1}{\left| I \left( \theta_b \right) \right|}\left[ 1 - \frac{\xi^2_0}{\varepsilon_r} - \cos \theta_b \right] \geq 2 \frac{\gamma_0}{p^2_0} \frac{E_0}{\varepsilon_r \tau_0}
. \label{eq:TrapCrit2}
\end{equation}
A spatially dependent dimensionless parameter quantifying the importance of collisionless detrapping is then given by the right hand side of Eq. (\ref{eq:TrapCrit2})
\begin{equation}
\Phi \equiv \frac{\gamma_0}{p^2_0} \frac{E_0}{\varepsilon_r \tau_0} = \left( \frac{\gamma_0}{p^2_0} \right) \left( \frac{qR_0}{a} \right) \left( \frac{1}{\varepsilon_r} \right) \left( \frac{E_1}{E_c}\right)  \left( \frac{a}{\tau_c c} \right)
. \label{eq:Phi}
\end{equation}
Taking $p_0$ to be the critical momentum to run away $p_c \sim 1/\sqrt{E_0}$, yields
\begin{equation}
\Phi_c \equiv \left( \frac{1}{b_g} \right) \left( \frac{qR_0}{a} \right) \left( \frac{1}{\varepsilon_r} \right) \left( \frac{E_1}{E_c}\right)^2  \left( \frac{a}{\tau_c c} \right)
. \label{eq:Phi2}
\end{equation}
This parameter quantifies the relative strength of the mirror force compared to electric field acceleration at $p_0=p_c$, where a value of order one indicates the inductive electric field will strongly impact electron detrapping.

To quantify how the above collisionless detrapping mechanism impacts RE generation, we first numerically evaluate the runaway probability function (RPF)~\cite{karney1986current} for different electric field strengths including both drag and pitch-angle scattering. Such a function is defined as the probability of an electron reaching the upper momentum boundary (taken to be $10p_c$) rather than slowing down to the low energy bulk, taken to be $\approx p_c \sqrt{ 1-\varepsilon_r }$. The RPF at $\theta=0$ is shown in Fig. \ref{fig:RPF1} for four different electric fields. These are computed by a direct solution to Eq. (\ref{eq:PF1}) with the Monte Carlo code JONTA~\cite{mcdevitt2026adjoint}.
For the case of a modest electric field $E_1=100$ [Fig. \ref{fig:RPF1}(a)], the RPF is largely hollowed out in the trapped-passing boundary located at $\xi_t \equiv \pm \sqrt{2\varepsilon_r / \left( 1+\varepsilon_r\right)} \approx \pm 0.65$ due to the inductive electric field averaging out over a complete bounce orbit.
As the electric field is increased [Fig. \ref{fig:RPF1}(b-d)], the trapped region begins to fill in, until at sufficiently large electric fields the RPF returns to a form familiar from slab RE calculations [Fig. \ref{fig:RPF1}(d)]~\cite{Liu:2016, zhang2017backward, McDevitt:hottail:2023}, with the momentum space structure of the RPF largely insensitive to the magnetic field geometry. 

To identify the mechanism underlying this transition, it will be useful to remove pitch-angle scattering, but still including collisional drag, such that the electrons' orbits are deterministic and detrapping by collisional scattering is no longer possible. Such a calculation will yield a binary result, either zero or one, rather than a probability distribution and hence will be referred to as a fate map. Fate maps corresponding to the RPFs in Figs. \ref{fig:RPF1}(a-d) are shown in Fig. \ref{fig:RPF1}(e-h). Here, it is apparent that for a modest electric field of $E_1/E_c = 100$, the fate map vanishes in the deeply trapped region, with complex oscillatory features along the trapped-passing boundary and in the counter passing region with $\xi > \left| \xi_t \right|$. Such structures result from electrons slowing down in a finite number of poloidal transits together with the $1/R$ dependence of the inductive electric field. A more detailed description of this subtle feature is given in Appendix \ref{sec:EMNM}. The qualitative difference between the fate map and RPF for $E_1=100$ [Fig. \ref{fig:RPF1}(e) versus \ref{fig:RPF1}(a)] is due to the absence of pitch-angle scattering in the former, such that deeply trapped electrons have no means of escaping the magnetic well.
In contrast, as the electric field is increased, the phase space inside the trapped region begins to fill with values of one [Fig. \ref{fig:RPF1}(f)], until the fate map is no longer sensitive to the magnetic geometry [Fig. \ref{fig:RPF1}(h)]. This indicates that the magnetic well is completely removed by the strong inductive electric field. At intermediate inductive electric fields, a complex series of lobes form in the previously trapped region, steadily eroding the impact of magnetic trapping.

Aside from providing a tool for diagnosing the physics of RE generation, the RPF can be used to evaluate the avalanche growth rate~\cite{Liu:2017, Arnaud2025arunaway}. An expression for the RE avalanche growth rate in tokamak geometry is given by Eq. (\ref{eq:REAV2}). Here, RP97's secondary birth curve $\xi_1 \equiv -\sqrt{\left( \gamma-1\right)/\left( \gamma+1\right)}$ will be localized inside the trapped region for $p=p_c$ and $E_1\gg1$, such that the strong modifications to the RPF evident in Fig. \ref{fig:RPF1} will sharply modify the efficiency of the RE avalanche mechanism. Our aim will be to identify a dimensionless parameter characterizing this transition. Previous work~\cite{mcdevitt2019runaway} postulated that the ratio of the bounce time to collisional detrapping rate $\nu_* = \tau_{bounce}/\tau_{dt}$, where $\tau^{-1}_{bounce} = \varepsilon^{1/2}_r v/\left( qR_0\right)$ and $\tau^{-1}_{dt}=\nu_D/\varepsilon_r$, evaluated at $p=p_c$ delineates the transition in the RPF, i.e.
\begin{equation}
\nu_{*,c} = \left( 1+Z_{eff} \right) \left( \frac{1}{b_g} \right)^2 \left( \frac{qR_0}{a} \right) \left( \frac{1}{\varepsilon_r} \right)^{3/2} \left( \frac{E_1}{E_c}\right)^2 \left( \frac{a}{\tau_c c}\right)
. \label{eq:nu1}
\end{equation}
Alternatively, collisionless detrapping based on a tilted magnetic well provides an additional candidate. 
For this latter process, the dimensionless parameter defined in Eq. (\ref{eq:Phi2}) provides a natural parameter. To distinguish between these two physical mechanisms, 
we define the parameter $S$:
\begin{equation}
S \equiv \frac{ G\left( \varepsilon_r \right) - \gamma_{av} \left( r\right) / \gamma_{av} \left( 0\right)}{G\left( \varepsilon_r \right) - \gamma_{ba}\left( r\right) / \gamma_{ba}\left( 0\right)}
, \label{eq:gamma2}
\end{equation}
where $G\left( \varepsilon_r \right)=1/\sqrt{1-\varepsilon^2_r}$ is an upper bound on the ratio of $\gamma_{av} (r) / \gamma_{av} (0)$ for a given $\varepsilon_r$ [see Appendix \ref{sec:MRE}], $\gamma_{av}$ is computed from the drift kinetic solver JONTA, $\gamma_{ba}$ is the bounce-averaged limit evaluated by a continuum relativistic Fokker-Planck solver, and we note $\gamma_{av} (0) = \gamma_{ba} (0)$. For $S=1$ this implies $\gamma_{av} (r) = \gamma_{ba} (r)$, whereas $S=0$ implies $\gamma_{av} (r) = \gamma_{av} (0) G\left( \varepsilon_r \right)$. 
By computing both $\gamma_{av}$ and $\gamma_{ba}$ at radial locations $r/a\in \left( 0, 1\right)$ and across a broad range of the parameters entering Eq. (\ref{eq:PF1}) $\left( \tau_0, E_0, q, \varepsilon \right)$ (whose ranges are given in the Appendix \ref{sec:JDS}) we empirically evaluate $S$ and thus determine if a criterion based on a single composite parameter can be identified.


\begin{figure}
\centering
\subfigure[]{\includegraphics[scale=0.45]{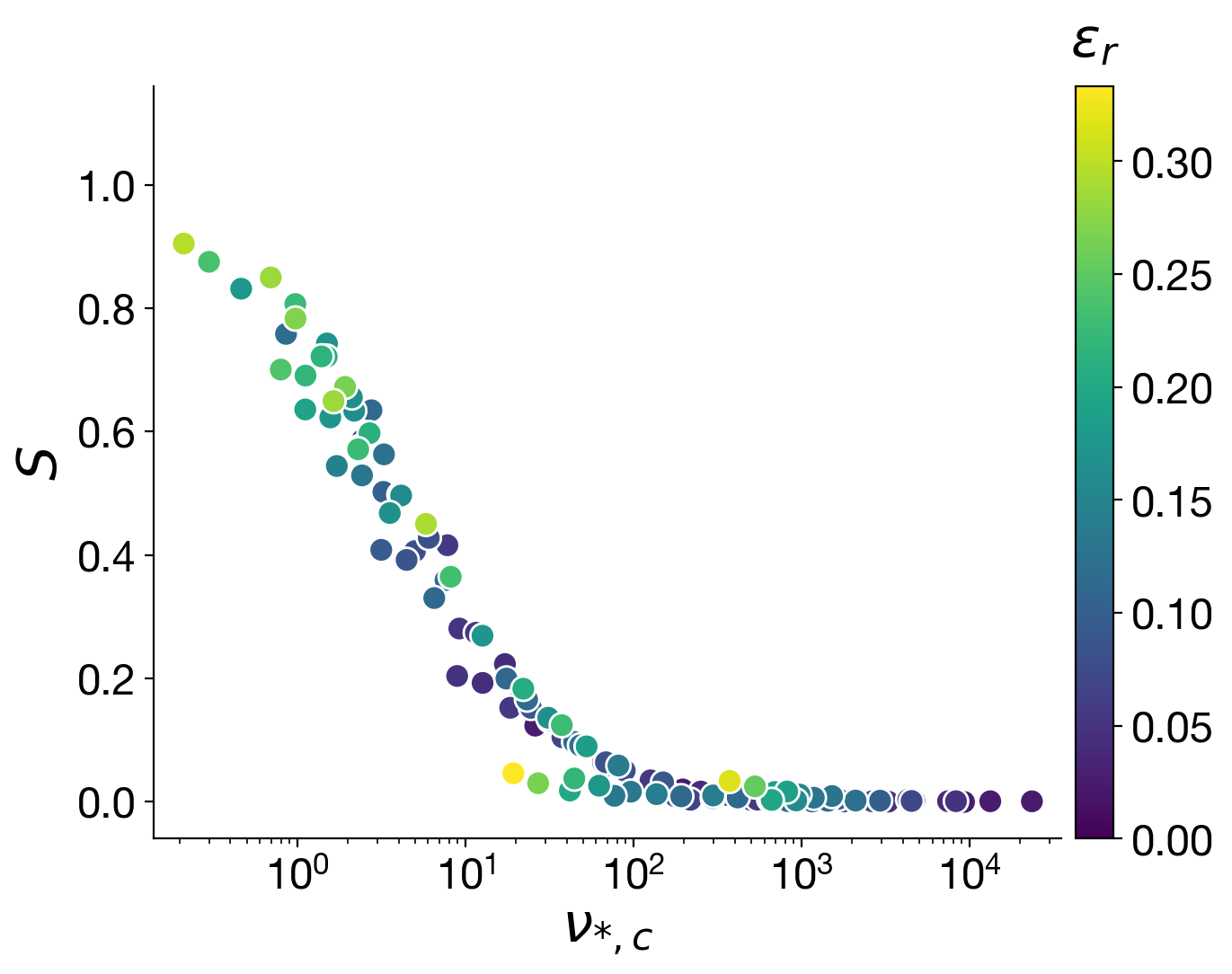}}%
\subfigure[]{\includegraphics[scale=0.45]{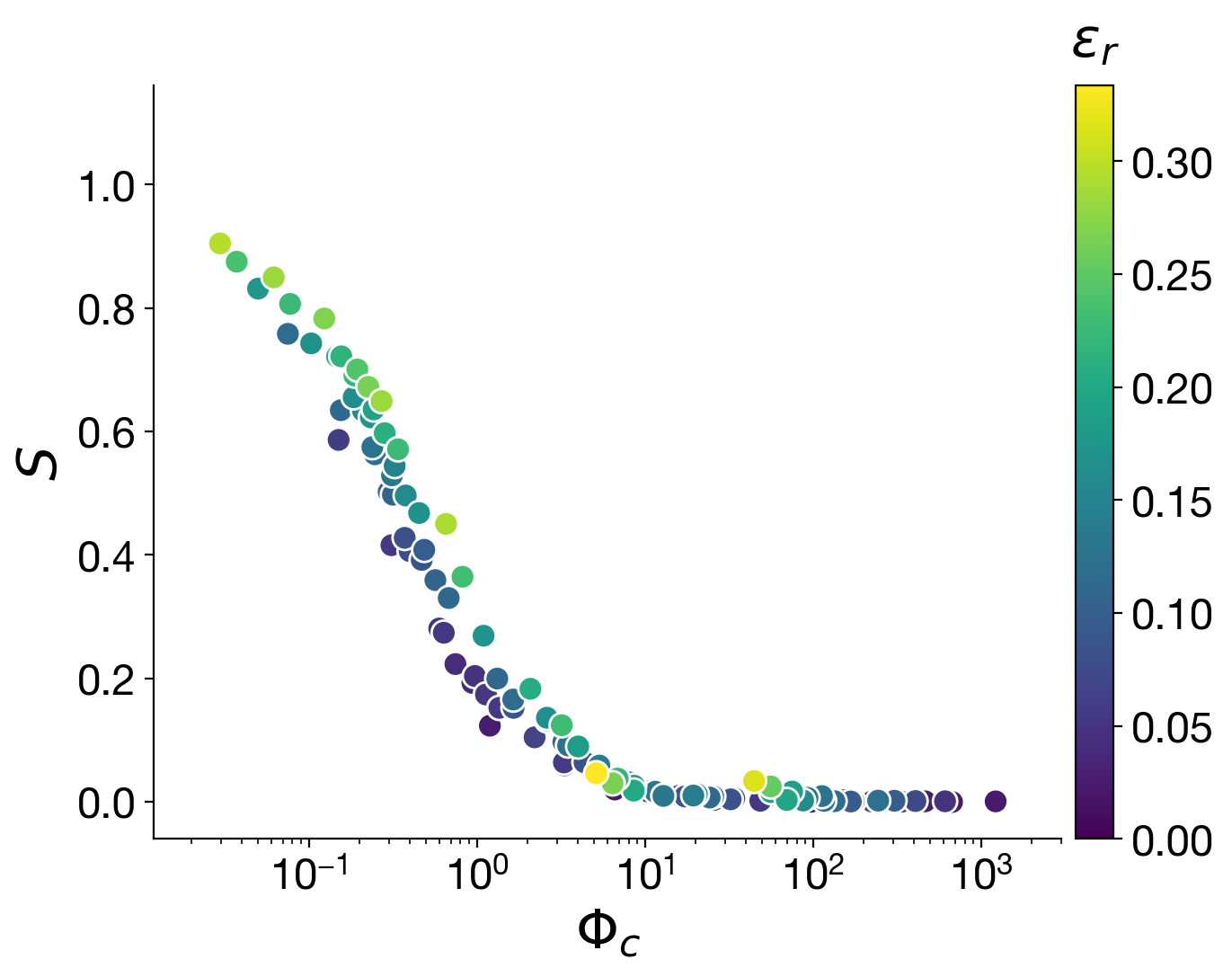}}
\subfigure[]{\includegraphics[scale=0.45]{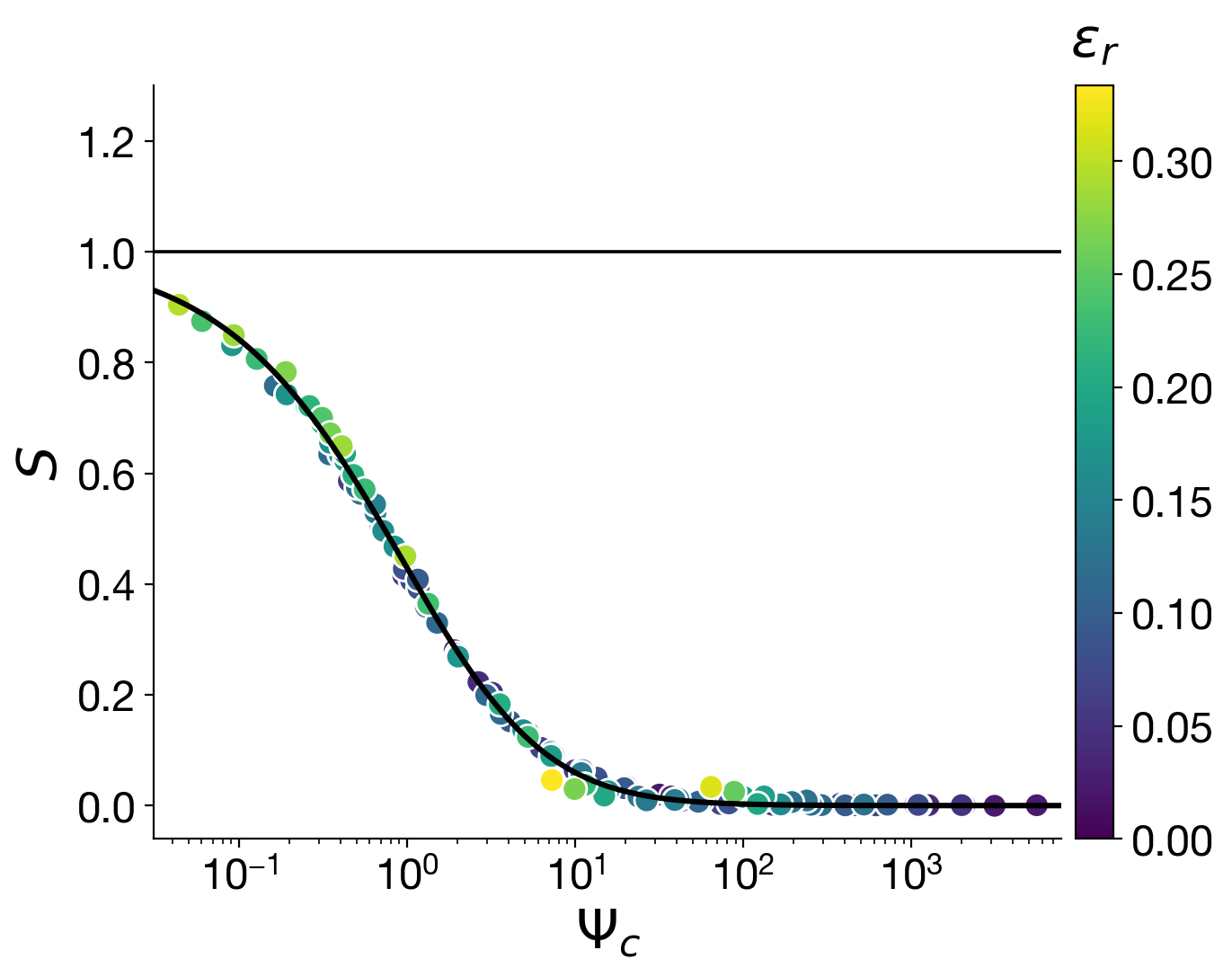}}%
\subfigure[]{\includegraphics[scale=0.45]{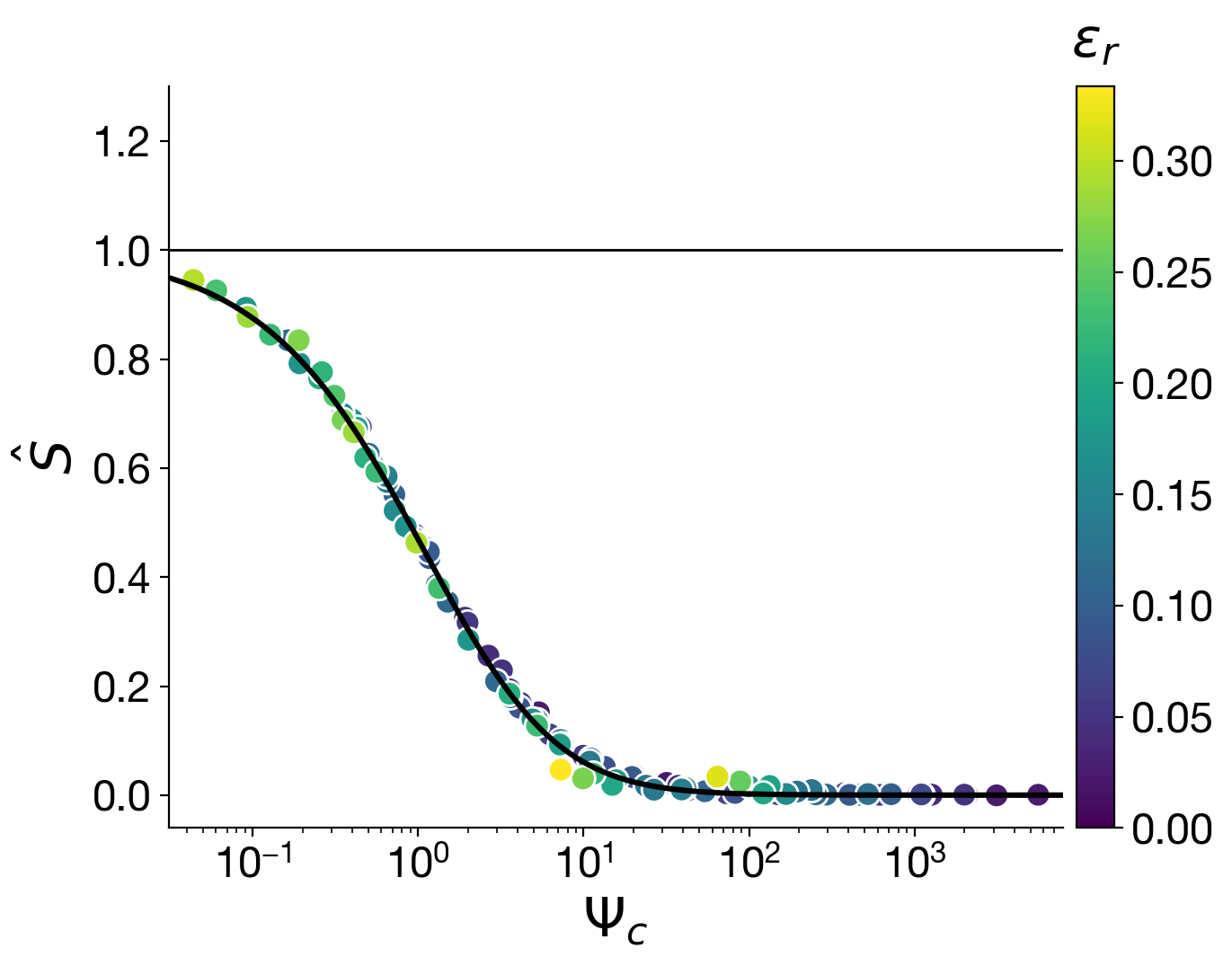}}
\caption{\label{fig:Phi1} The factor $S$ defined by Eq. (\ref{eq:gamma2}) evaluated across the JONTA data. Panel (c) includes a two parameter fit of the $S = \left[ 1+\left( \Psi_c/\Psi_0\right)^a\right]^{-2}$ curve to JONTA data. Panel (d) includes an analogous fit using the $\gamma_{RP}$ instead of $\gamma_{ba}$ in Eq. (\ref{eq:gamma2}).
}
\end{figure}

The values of $S$ are plotted against the dimensionless quantities defined by Eqs. (\ref{eq:Phi2}) and (\ref{eq:nu1}) in Figs. \ref{fig:Phi1}(b) and \ref{fig:Phi1}(a), respectively. Both quantities organize the JONTA data well, but significant scatter remains. A careful treatment of the data reveals a far weaker dependence on $Z_{eff}$ than the $\left( 1+ Z_{eff}\right)$ predicted by $\nu_{*,c}$, implying that pitch-angle scattering is not the dominant detrapping process. $\Phi_c$ on the other hand, provides a modest improvement, but does not fully capture the inverse aspect ratio dependence $\varepsilon_r$. This is evident in Fig. \ref{fig:Phi1}(b), where light color markers, which have larger values of $\varepsilon_r$, are systematically to the right of darker markers with smaller $\varepsilon_r$. It is interesting to note, however, that $\nu_{*,c}$ correctly captures the $\varepsilon_r$ dependence of the data, where it scales with $1/\varepsilon^{3/2}_r$ instead of a factor of $1/\varepsilon_r$ in $\Phi_c$. Noting the location of the trapped-passing boundary has the required scaling $\xi_t =\sqrt{2\varepsilon_r/\left( 1 + \varepsilon_r \right)}$, we define an additional dimensionless quantity of the form 
\begin{equation}
\Psi_c \equiv \Phi_c/ \xi_t
. \label{eq:Phi3}
\end{equation}
Figure \ref{fig:Phi1}(c) shows a comparison of $\Psi_c$ with the JONTA data set. It is evident that the data collapses onto a single curve with small and large values of $\varepsilon_r$ now merged. Such a hybrid parameter suggests that no single detrapping mechanism solely sets the transition, but instead distinct mechanisms work in synergy.


The identification of a single composite parameter capturing the transition in the RPF structure enables a generalization of bounce-averaged or analytic formulations of RE generation. This can be done by fitting a one-dimensional function to the collapsed $S$ function. We find an excellent fit [Fig. \ref{fig:Phi1}(c)] with a function of the form $S \left( \Psi_c \right) = \left[ 1+\left( \Psi_c/\Psi_0\right)^a \right]^{-2}$ containing only two fit factors whose values are found to be $a=0.7684$ and $\Psi_0=2.3024$. 
If we now solve Eq. (\ref{eq:gamma2}) for $\gamma_{av}$, and note $\gamma_{av} \left( 0 \right) = \gamma_{ba} \left( 0 \right)$, this yields
\begin{equation}
\gamma_{av} (r) = \gamma_{ba} (0) \left[ G\left( \varepsilon_r \right) - \left( G\left( \varepsilon_r \right) - \frac{\gamma_{ba}\left( r\right)}{\gamma_{ba} \left( 0\right)}\right) S \left( \Psi_c \right)\right]
, \label{eq:gamma3}
\end{equation}
this provides a means of generalizing bounce-averaged predictions to the drift kinetic limit.

\begin{figure}
\centering
\subfigure[]{\includegraphics[scale=0.5]{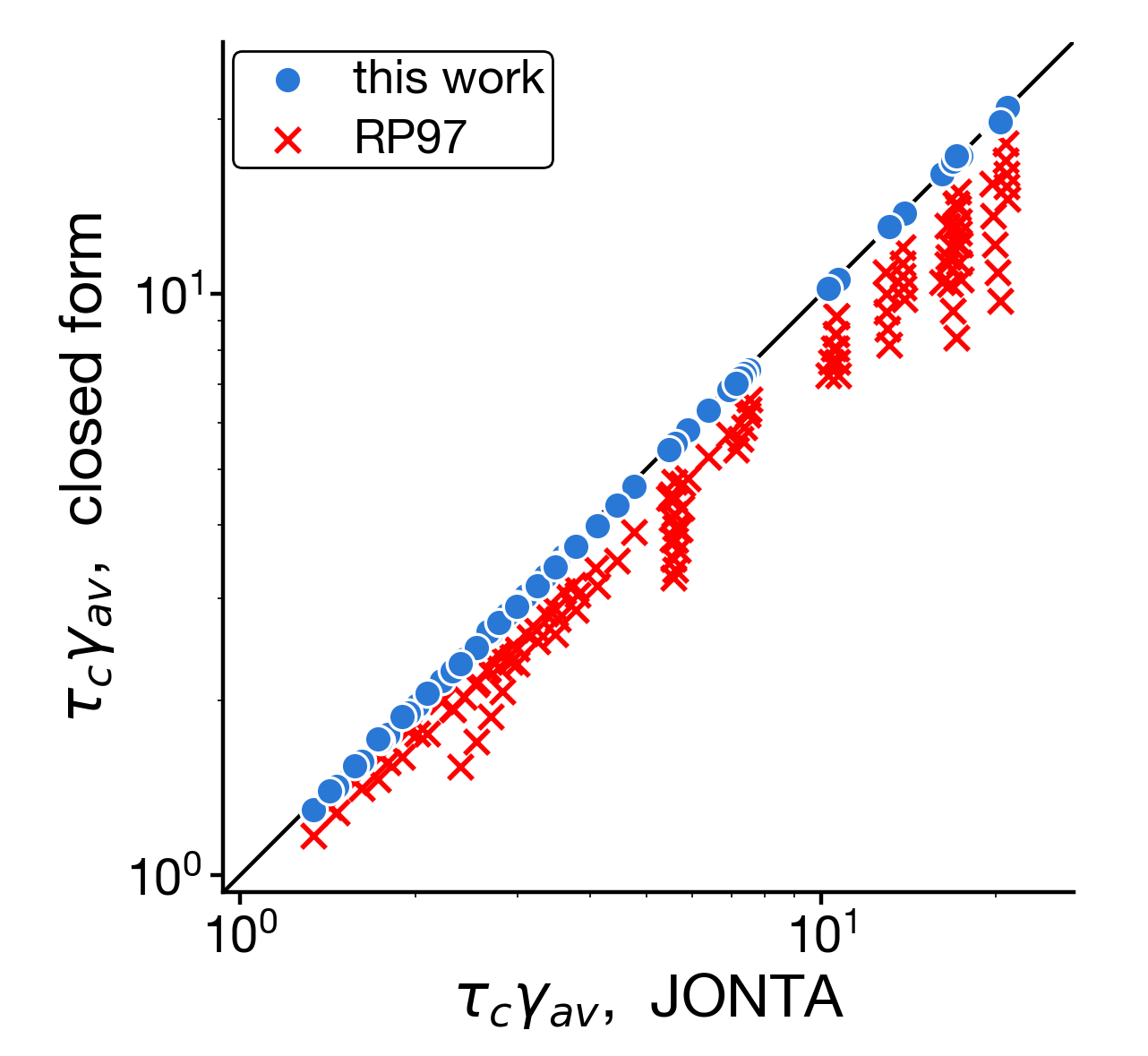}}%
\subfigure[]{\includegraphics[scale=0.5]{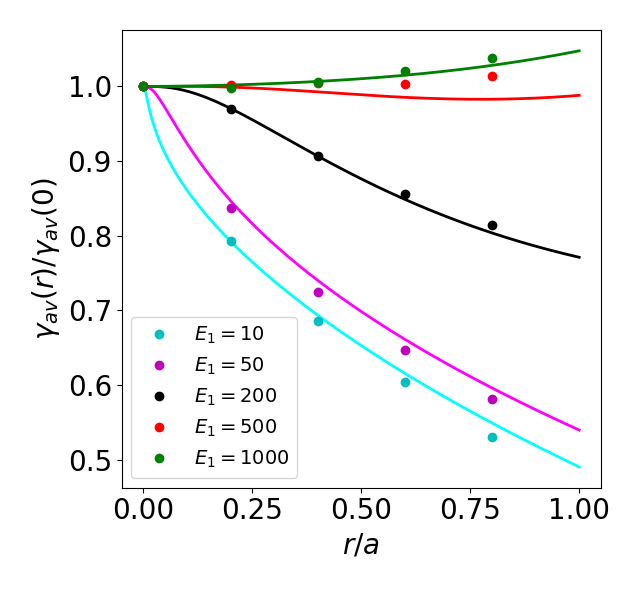}}
\caption{\label{fig:RP1}(a) Comparison of the avalanche growth rate predicted by Eq. (\ref{eq:RP1}) with JONTA data. The markers indicate the RP97 formula. (b) Normalized RE avalanche growth rate versus radius for different electric field strengths. The solid curves indicate Eq. (\ref{eq:RP1}) and the markers taken from Fig. 6 of Ref. \cite{mcdevitt2019runaway}. The electric field and radius are varied as indicated, and the other parameters are taken to be $\left(c\tau_c/a\right)=2.505\times 10^6$, $a/R_0=1/3$, $Z_{eff}=1$, $\ln \Lambda = 10$, and $q\left( r\right) = 2.1 + 2 \left( r/a\right)^2$.}
\end{figure}

A generalization of the RP97 growth rate formula can be obtained by replacing $\gamma_{ba}$ in Eq. (\ref{eq:gamma2}) with the analytic expression $\gamma_{RP}$ given by Eq. (18) of RP97~\cite{Rosenbluth:1997}, whose form is given by Eqs. (\ref{eq:RP1sub1}) and (\ref{eq:RP2}) of Appendix \ref{sec:RPF}. We note that the integral form of the effective passing fraction is used in this expression, rather than its small inverse aspect ratio approximation.
The new $S$, which we denote by $\hat{S}$, will then act as a switch between $\gamma_{RP}(0) G \left( \varepsilon_r\right)$ and $\gamma_{RP}(r)$. The fitted curve for $\hat{S}$ is shown in Fig. \ref{fig:Phi1}(d), 
with fit parameters $\hat{\Psi}_0=2.5710$ and $\hat{a}=0.8237$, yielding the generalized RP97 avalanche growth rate formula
\begin{equation}
\gamma_{av} (r) = \gamma_{RP} (0) \left[ G\left( \varepsilon_r \right) - \left( G\left( \varepsilon_r \right) - \frac{\gamma_{RP}\left( r\right)}{\gamma_{RP} \left( 0\right)}\right) \hat{S} \left( \Psi_c \right)\right]
, \label{eq:RP1}
\end{equation}
where we took $\gamma_{av} (0) \approx \gamma_{RP} (0)$. The accuracy of this generalized avalanche growth rate formula is quantified in Fig. \ref{fig:RP1}. The root-mean-square (RMS) with direct drift kinetic JONTA simulations is 1.79\% with a worst case of 3.4\%. This compares with an RMS of 25\% and worst case of 52\% from the original RP97 expression. 

In this paper we have derived a criterion for a novel means of electron detrapping that emerges through the tilting of the magnetic well by a strong inductive electric field. Such a modification to the trapped-passing boundary is distinct from the well established modification that emerges due to an electrostatic potential~\cite{indireshkumar1993effect, shaing1994resonance, chang1989tokamak}. Specifically, while an electrostatic field modifies the energy invariant $m_e c^2 \left( \gamma - 1\right) - e\phi \left( \theta \right)$ leading to a shift or deepening of a magnetic well, an inductive electric field $E_\varphi = -\partial A_\varphi / \partial t$ removes energy as an invariant, due to the explicit time dependence in the Hamiltonian. This results in the distinctive tilting of the magnetic well evident in Fig. \ref{fig:TiltWell}. A composite quantity $\Psi_c$, directly linked to the strength of the tilt in the magnetic well is shown to collapse the four parameters $\left( E_0, \tau_0, \varepsilon_r, Z_{eff}\right)$ characterizing the solution of the drift kinetic equation onto a single one-dimensional curve. By performing a two-parameter fit to this curve a generalization of the RP97 theory to the drift kinetic regime is achieved, with maximum deviations of less than $4\%$ when well above threshold. While this paper has employed the relatively simple collision coefficients appropriate in the completely screened limit, noting that the tilted well mechanism is fundamentally collisionless, we expect it to be active in plasmas with large quantities of partially screened impurities~\cite{Hesslow:2017}, where Ref. \cite{arnaud2024impact} has already documented large deviations from bounce-averaged results for partially ionized plasmas under disruption conditions.

\section*{AI disclosure} 

During the preparation of this work the authors used Claude Opus 4.8, 5 and 5.5 to assist with the development, testing and verification of the Monte Carlo and analysis codes used to produce the results reported here; with the numerical analysis and fitting of the resulting data sets; and with literature searching. Aside from this paragraph, the manuscript text was written entirely by the authors. The figures were produced by the authors using analysis scripts developed with the assistance of these tools. The authors reviewed and verified all outputs and take full responsibility for the content of this paper.

\begin{acknowledgements}

This work was supported by the Department of Energy, Office of Fusion Energy Sciences under awards DE-SC0024634 and DE-SC0024649. The authors acknowledge the University of Florida Research Computing for providing computational resources that have contributed to the research results reported in this publication. This work was supported by the U.S. Department of Energy through the Los Alamos National Laboratory. Los Alamos National Laboratory is operated by Triad National Security, LLC, for the National Nuclear Security Administration of U.S. Department of Energy (Contract No. 89233218CNA000001).

\end{acknowledgements}


\appendix

\section{Emergence of a non-monotonic runaway probability function}
\label{sec:EMNM}

\begin{figure}
\centering
\subfigure[]{\includegraphics[scale=0.55]{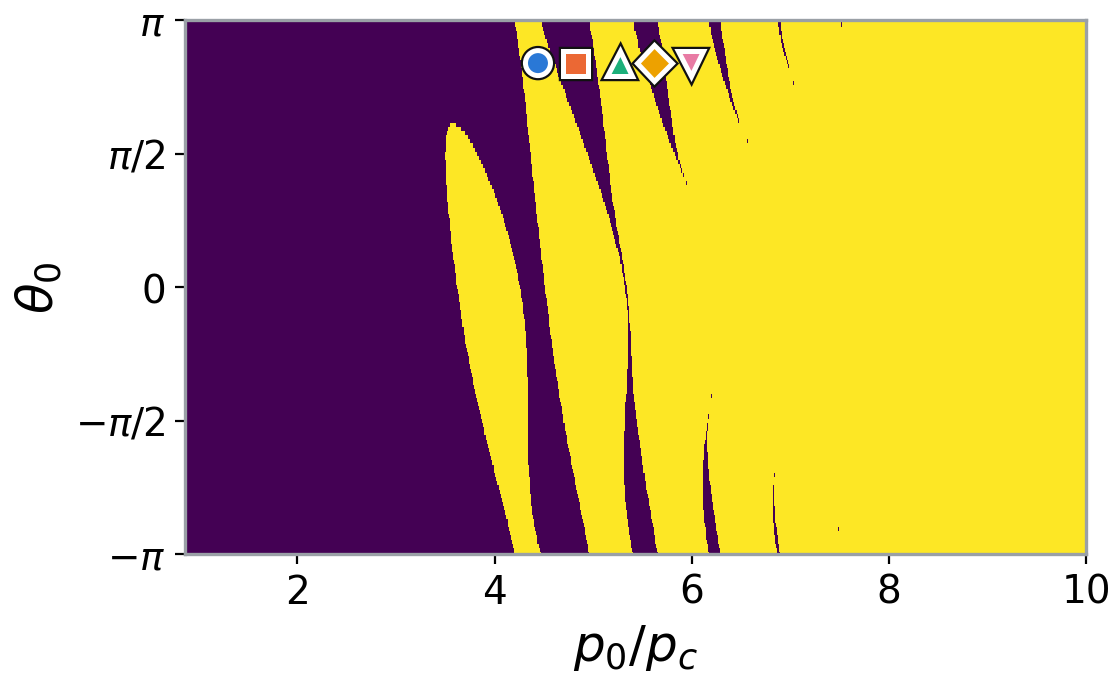}}%
\subfigure[]{\includegraphics[scale=0.55]{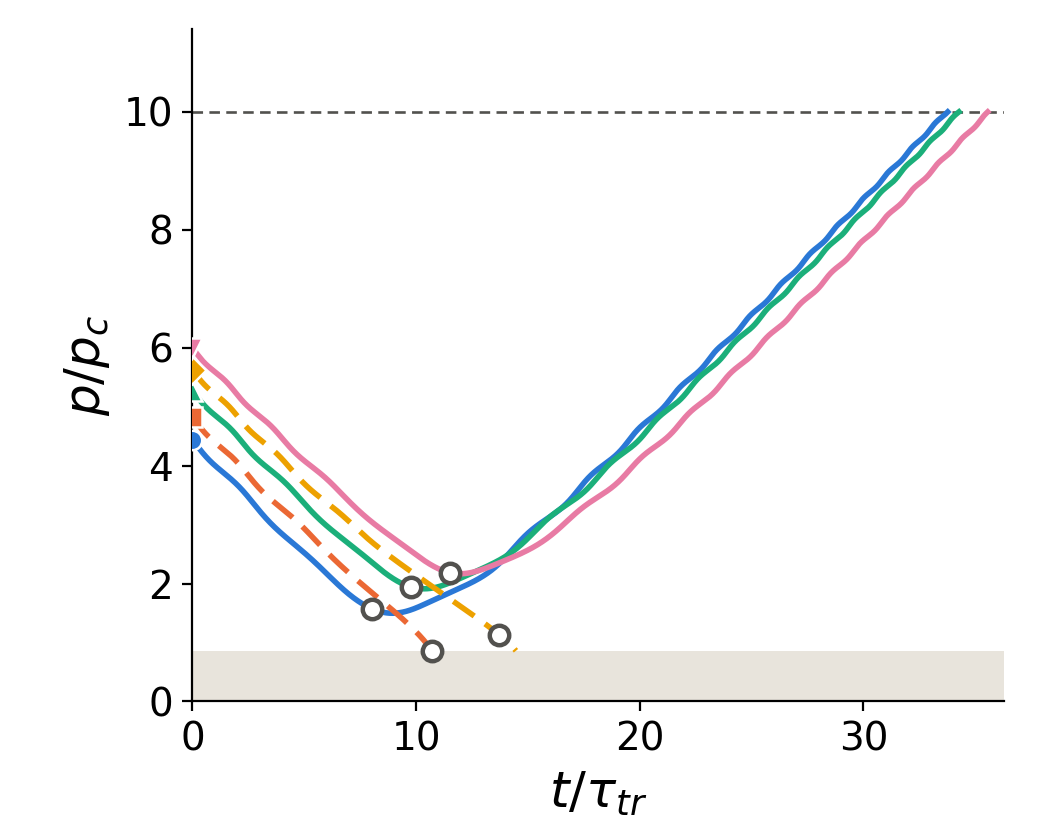}}
\subfigure[]{\includegraphics[scale=0.55]{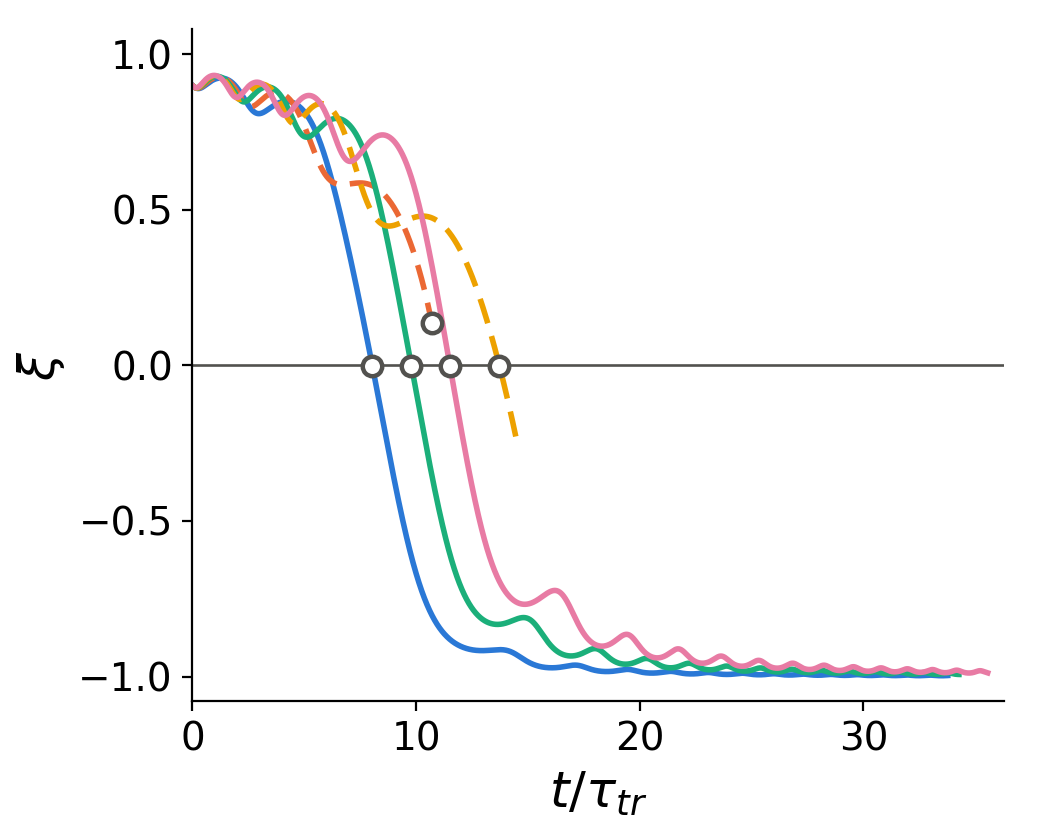}}%
\subfigure[]{\includegraphics[scale=0.55]{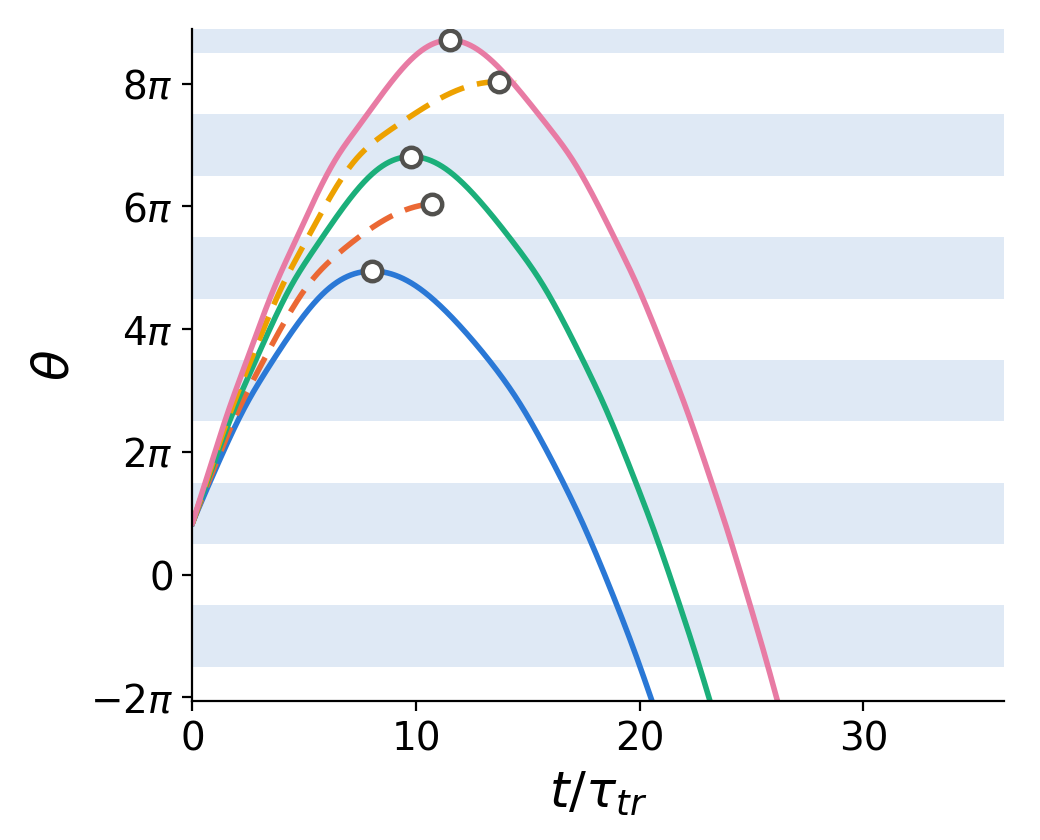}}
\caption{\label{fig:RPF2} (a) Deterministic fate map for electrons with $\xi=0.9$ at different initial momentum $p$. Yellow regions run away, purple regions do not. Markers indicate initial locations of the electron orbits shown in the neighboring panels. Panels (b), (c) and (d) show momentum, pitch, and poloidal orbits of example electron orbits. Dashed curves indicate electrons lost to the bulk with solid curves indicating those that run away. The circle markers illustrate when an electron crosses $\xi=0$. The parameters are the same as Figs. \ref{fig:RPF1}(c) and (g). Time is in units of poloidal transits for electrons with $p=p_{max}$.}
\end{figure}

A unique feature of the strong field regime of RE generation is that the RPF no longer monotonically increases with energy at all phase space locations. To understand these non-monotonic features it will be useful to consider the fate map in the $\theta$-$p$ plane for counter propagating electrons (i.e. $\xi =0.9 > \left| \xi_t \right|$), where we will consider several deterministic electron orbits with the same initial pitch and poloidal angle, but different momenta $p$ [indicated by markers in Fig. \ref{fig:RPF2}(a)]. This corresponds to a fate map of the RPF shown in Fig. \ref{fig:RPF1}(c).
Beginning with the lowest energy electron considered (blue curves), this electron makes a full poloidal transit, but is reflected by a combination of the electric field and mirror force in Eq. (\ref{eq:PF1b}) in the strong field region preventing it from completing a second poloidal transit. Once turned, the electron is then accelerated to relativistic energies by the inductive electric field. Conceptually, the reflection of the electron for $\xi > 0$, but then its subsequent acceleration for $\xi < 0$, is due to the tilted well structure evident in Fig. \ref{fig:TiltWell}.
Considering now the red orbit with larger $p$, due to its higher energy and the $1/p$ scaling of the electric field term in Eq. (\ref{eq:PF1b}), it is able to overcome the tilted well on its second poloidal transit ($\theta = 5\pi$), but does not have sufficient energy to make it back to the strong field side. As the electron continues to propagate with $\xi > 0$, drag together with the electric field slow the electron until it returns to the bulk on the weak field side. An electron with slightly higher momentum (green orbit) is able to make it past the weak field region, after which it is reflected and runs away.
This alternating pattern continues, giving rise to a fate map that alternates between zero and one as the electron's energy is increased. The inclusion of pitch-angle scattering would smear the fate map shown in Fig. \ref{fig:RPF2}(a) into a probability function, but the non-monotonic character remains if the number of poloidal transits taken before running away or returning to the bulk is modest.

\section{Avalanche growth rate formula in toroidal geometry} 
\label{sec:AGR}

\begin{figure}
\centering
\subfigure[]{\includegraphics[scale=0.5]{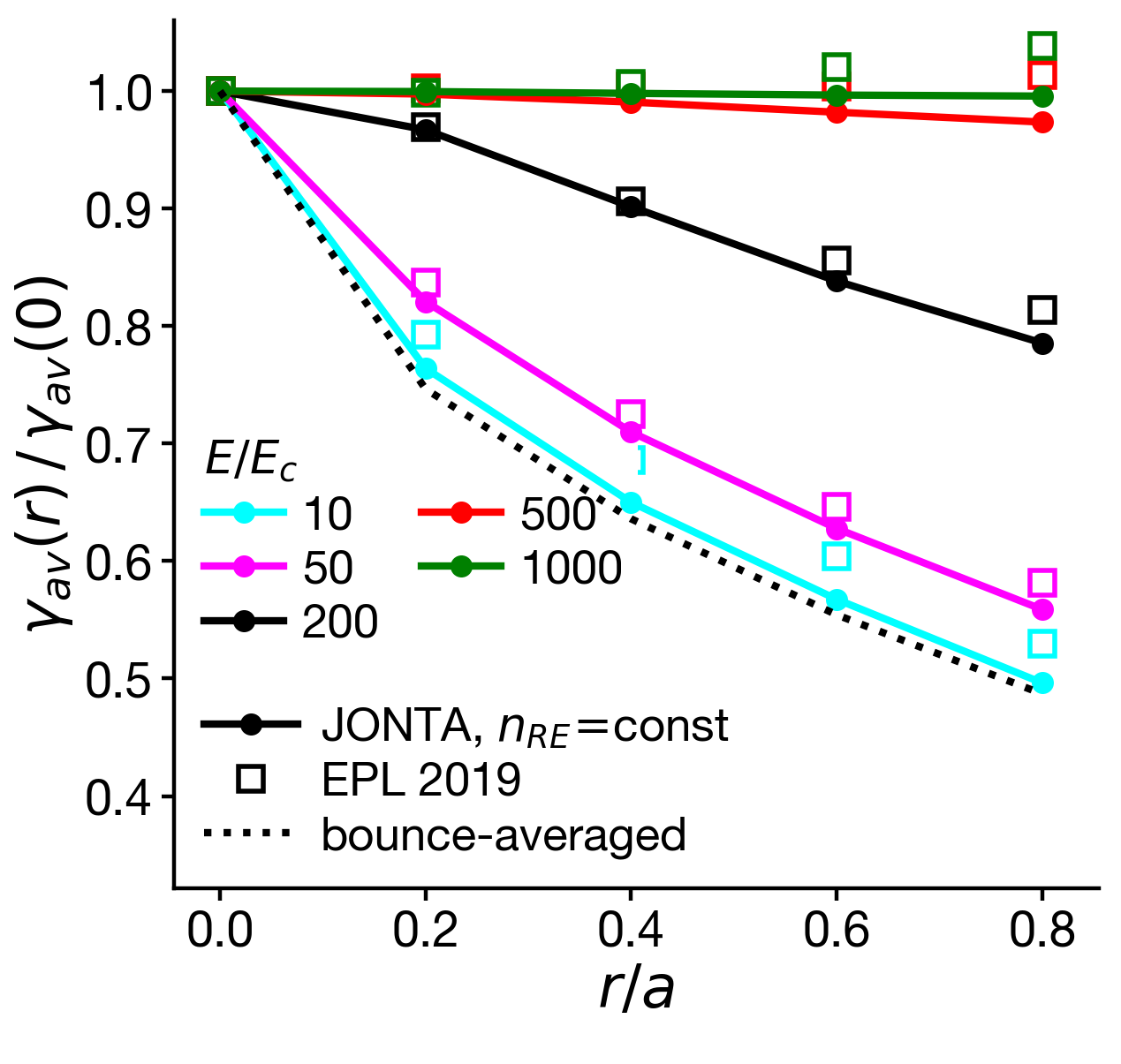}}
\subfigure[]{\includegraphics[scale=0.5]{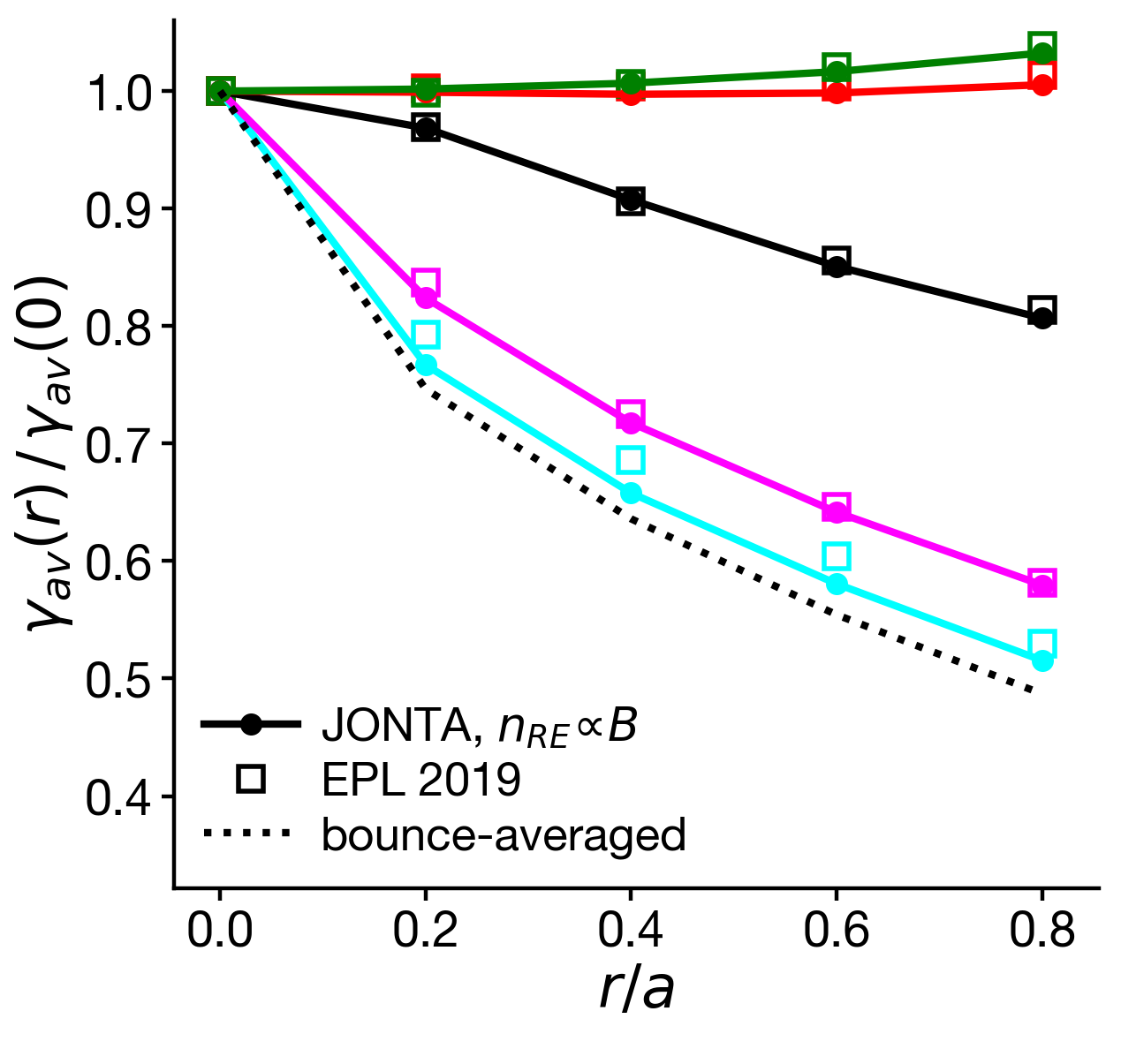}}
\caption{\label{fig:REAV1} Comparison of Eq. (\ref{eq:REAV2}) using an RPF computed from JONTA with Fig. 6 of Ref.~\cite{mcdevitt2019runaway}. Panel (a) is for $n_{RE}=\text{const}$ whereas panel (b) is for $n_{RE} \left( \theta\right) \propto B \left( \theta\right)$. The electric field and radius are varied as indicated, and the other parameters are taken to be $\left( c\tau_c/a \right)=2.505\times 10^6$, $a/R_0=1/3$, $Z_{eff}=1$, $\ln \Lambda = 10$, and $q\left( r\right) = 2.1 + 2 \left( r/a\right)^2$. The upper momentum was set to $p_{max}=30p_c$.}
\end{figure}

Secondary electrons are generated at a rate~\cite{Liu:2017}
\begin{equation}
\left. \frac{d n_{RE}}{dt} \right|_{av} = \int d^3 p S \left( p, \xi, \theta \right) P \left( p, \xi,\theta \right)
, \label{eq:REAV1}
\end{equation}
where $n_{RE}$ is the RE density, $S\left(p, \xi, \theta \right) \equiv \int d^3 p^\prime S_0 \left(p^\prime , \xi^\prime , p , \xi \right) f_{RE} \left( p^\prime, \xi^\prime, \theta \right)$, $f_{RE}$ is the electron distribution, and $S_0 \left(p^\prime , \xi^\prime , p , \xi \right)$ is set by the M\o ller source~\cite{Moller:1932, Boozer:2015}. Employing the RP97 approximation to the RE distribution, $p\to \infty$ and $\xi = -1$, the avalanche growth rate can be written as~\cite{arnaud2024physics}:
\begin{align}
\tau_c \gamma_{av} & \equiv \frac{\tau_c }{\left\langle n_{RE} \right\rangle_{FS}} \left\langle \left. \frac{d n_{RE}}{dt} \right|_{av} \right\rangle_{FS} = \frac{1}{2\ln \Lambda} \left\langle \frac{n_{RE} \left( \theta \right)}{\left\langle n_{RE} \right\rangle_{FS}}  \int dp \frac{v}{\left( \gamma - 1\right)^2}  P \left( p, \xi_1, \theta \right) \right\rangle_{FS}
, \label{eq:REAV2}
\end{align}
where $\left\langle \cdots \right\rangle_{FS} = \left( \oint \frac{d\theta }{\mathbf{B} \cdot \bm{\nabla} \theta }\right)^{-1} \oint \frac{d\theta }{\mathbf{B} \cdot \bm{\nabla} \theta} \left( \cdots\right)$ is a flux surface average and $\xi_1 = -\sqrt{\left( \gamma-1\right)/\left( \gamma+1\right)}$ is RP97's secondary electron birth curve. For the high energies typical of primary REs, the particle distribution will be to lowest order a function of its collisionless invariants, $f_{RE} = f_{RE} \left( p, \lambda, \psi \right)$, where $\lambda \equiv \left( 1-\xi^2\right)/B\left( \theta\right)$. Noting $n_{RE} \equiv \int d^3p f_{RE} \left( p, \lambda, \psi \right)$, a short calculation yields $n_{RE} \left( \theta \right) \propto B \left( \theta\right)$ for REs with $\left| \xi \right| \approx 1$. This proportionality to the magnetic field strength, equivalent to $j_{RE}/B \approx -ec n_{RE}/B = \text{const}$, leads to a slight increase in the avalanche growth rate since more secondary electrons are generated on the inboard side where the electric field is larger. This is shown in Fig. \ref{fig:REAV1}, where the avalanche growth rate is compared with results from Ref. \cite{mcdevitt2019runaway}, which employed the RAMc drift kinetic RE solver with a full M\o ller source in tokamak geometry. Here, the avalanche growth rate is computed for both $n_{RE} = \text{const}$ and $n_{RE} \propto B \left( \theta \right)$. It is evident that $n_{RE} \propto B$ gives a systematically more accurate result, and recovers the slight increase in the RE growth rate at large radii for the largest electric fields considered. This manuscript will thus use the more accurate $n_{RE} \propto B$ throughout.

\section{Maximum Runaway Electron Enhancement Factor} 
\label{sec:MRE}

An upper bound on the increase of the avalanche growth rate at large radii can be computed in the asymptotically large electric field limit. In this limit, secondaries are accelerated by the inductive electric field before moving an appreciable distance poloidally, hence their generation rate is set by the local inductive electric field $E_\varphi \propto 1/R$. 
Noting that the dominant contribution to the RE avalanche is for electrons near $p_c \sim 1/\sqrt{E_\varphi-1} \approx 1/\sqrt{E_\varphi} = 1/\sqrt{E_1 R_0/R}$, we can introduce the new momentum variable $u \equiv p/p_c = p \sqrt{E_1R_0/R}$. For sufficiently large electric field, the RPF will only depend on the poloidal angle through the local electric field strength, which is now captured by the normalized momentum variable $u$, allowing $P \left( p, \xi_1, \theta \right) \approx P \left( u, \xi_1\right)$. Further noting the weight in the momentum integral in Eq. (\ref{eq:REAV2}) can be approximated as $v/\left( \gamma - 1 \right)^2 \approx 4/p^3$ for $p \ll 1$, and using $u$ as the momentum space integration variable, allows Eq. (\ref{eq:REAV2}) to be written as
\begin{align}
\tau_c \gamma_{av} & = \frac{1}{2\ln \Lambda} \left\langle \frac{n_{RE} \left( \theta \right)}{\left\langle n_{RE} \right\rangle_{FS}}  \int dp \frac{4}{p^3}  P \left( u, \xi_1 \right) \right\rangle_{FS} =\left\langle \frac{n_{RE} \left( \theta \right)}{\left\langle n_{RE} \right\rangle_{FS}} \frac{R_0}{R} \right\rangle_{FS} \gamma_{av0}
, \label{eq:REAV3}
\end{align}
where $\tau_c\gamma_{av0} \equiv \left( 2E_1/\ln \Lambda \right) \int du P  \left( u, \xi_1 \right) / u^3$ is the slab RE generation rate, which satisfies $\gamma_{av0} = \gamma_{av}(0)$. For $n_{RE} = \text{const}$, it is straightforward to show the flux surface averaged quantity on the second line of Eq. (\ref{eq:REAV3}) is unity. However, for $n_{RE} \propto B \propto 1/R$ we have
\begin{equation}
G \left( \varepsilon_r \right) \equiv \left\langle \frac{n_{RE} \left( \theta \right)}{\left\langle n_{RE} \right\rangle_{FS}} \frac{R_0}{R} \right\rangle_{FS} = \frac{1}{\sqrt{1-\varepsilon^2_r}}
. \label{eq:REAV4}
\end{equation}
Thus, an upper bound on the avalanche growth rate is given by $\gamma^{(max)}_{av} = \gamma_{av} (0) G \left( \varepsilon_r \right) = \gamma_{av} (0)/\sqrt{1-\varepsilon^2_r}$. 

\section{JONTA data set} 
\label{sec:JDS}

The Monte Carlo data generated for this paper spans the four-dimensional parameter space $\left( E_0, \tau_0, \varepsilon_r, Z_{eff}\right)$. The range of values for the data set are chosen such that $E_1/E_c \in \left[ 100,  1000\right]$,  $\left( c \tau_c /a \right) \in \left[ 1.8, 320\right] \times 10^{5}$, $Z_{eff} \in \left[ 1, 5\right]$, $q \in \left[ 1, 5\right]$, $r/a \in \left[ 0, 1\right]$, and inverse aspect ratios of $\varepsilon \equiv a/R_0 \in \left[1/10,1/3\right]$. Such a parameter regime captures regions substantially above threshold, where collisionless detrapping is expected to impact RE generation. A random sample of 144 parameters values was generated, of which 24 were on-axis, leaving 120 values for constraining the collapse parameter. For each of these 144 parameter values, 128 values along $\xi_1$ are chosen at 16 different poloidal locations. At each of these 294,912 points, 32,768 marker particles are evolved to provide an accurate estimate of the RPF at this location.

\section{Rosenbluth-Putvinski formula and neoclassical factor} 
\label{sec:RPF}

Equation (18) of RP97 can be written as~\cite{Rosenbluth:1997}
\begin{align}
\tau_c \gamma_{RP} = \frac{1}{\ln \Lambda} \sqrt{\frac{\pi \phi }{3 \left( Z_{eff}+5\right)}} \left( E_1 - 1\right) \left( 1 - \frac{1}{E_1} + \frac{4\pi \left( Z_{eff}+1\right)^2}{3\phi \left( Z_{eff} + 5\right) \left( E^2_1 +4/\phi^2 - 1\right)} \right)^{-1/2}
, \label{eq:RP1sub1}
\end{align}
where $\phi$ is the effective passing electron fraction, which was evaluated in the small inverse aspect ratio limit in RP97. Such an approximation is generally not well satisfied for inverse aspect ratios typical of tokamak plasmas. We will thus use the effective passing fraction as defined in Ref.~\cite{lin1995upper}, which is given by the integral
\begin{equation}
\phi\left(\varepsilon_r\right) = \frac{3}{4}\left\langle b^2\right\rangle_{FS}
\int_0^{1/b_{max}} \frac{\lambda\, d\lambda}{\left\langle \sqrt{1-\lambda b}\right\rangle_{FS}}
, \label{eq:RP2}
\end{equation}
where $b\left(\theta\right) = 1 / \left(1+\varepsilon_r\cos\theta \right)$ and $b_{max} = 1 / \left( 1-\varepsilon_r \right)$, and it can be verified that $\phi\left( 0 \right) = 1$, thus recovering the cylindrical limit. Such a form gives a far better approximation to the bounce-averaged limit compared to its small inverse aspect ratio approximation.

\bibliographystyle{apsrev}
\bibliography{../../../../bib_files/ref}

\end{document}